\documentclass[twocolumn,showpacs,amsmath,amssymb,prl]{revtex4}

\usepackage{graphicx}% Include figure files
\usepackage{dcolumn}% Align table columns on decimal point
\usepackage{bm}% bold math

\begin{document}
%\preprint{CERC}
\title{
Hall Effect of Light}
% Force line breaks with \\
%
\author{Masaru Onoda$^1$}
\email{m.onoda@aist.go.jp}
\author{Shuichi Murakami$^{2}$}
\email{murakami@appi.t.u-tokyo.ac.jp}
\author{Naoto Nagaosa$^{1,2}$}
\email{nagaosa@appi.t.u-tokyo.ac.jp}
\affiliation{
$^1$Correlated Electron Research Center (CERC),
National Institute of Advanced Industrial Science and Technology (AIST),
Tsukuba Central 4, Tsukuba 305-8562, Japan\\
$^2$CREST, Department of Applied Physics, University of Tokyo,
Bunkyo-ku, Tokyo 113-8656, Japan
}
%\affiliation{%
\date{\today}% It is always \today, today,
% %  but any date may be explicitly specified
%
% %  but any date may be explicitly specified
%
\begin{abstract}
We derive the semiclassical equation of motion for the wave-packet of light
taking into account the Berry curvature in momentum space.
This equation naturally describes the interplay between 
orbital and spin angular-momenta, i.e., 
the conservation of total angular-momentum of light. 
This leads to the shift of wave-packet motion perpendicular to 
the gradient of dielectric constant, i.e.,
the polarization-dependent Hall effect of light.
An enhancement of this effect in photonic crystals is also proposed. 
\end{abstract}
% % Classification Scheme.
%%\keywords{Suggested keywords}%Use showkeys class option if keyword
%  %display desired
\maketitle

The similarity between geometrical optics and particle dynamics
has been the guiding principle to develop the quantum mechanics
in its early stage. One can consider a trajectory or ray of light at
the length scale much larger than the wavelength $\lambda$. 
By setting  $\hbar = c = 1$, 
the equation for the eikonal $\psi$ in the geometrical optics reads
$
[\partial_{\mu} \psi(x)]^2 =
[\bm{\nabla}\phi(\bm{r})]^2 - n(\bm{r})^2 = 0
$,
where $x_\mu = (\bm{r}, t)$ is the four-dimensional coordinates, $\psi(x) = 
\phi(\bm{r}) - n(\bm{r})t$, and $n(\bm{r})$ is the refractive index
slowly varying within the wavelength \cite{Born}. 
This equation is identical to the Hamilton-Jacobi equation for 
the particle motion by replacing $\psi$ by the action $\mathcal{S}$. 
%It is well known that this analogy has been the guiding principle for 
%early development of quantum mechanics. 
The equation for the ray of light can be derived 
from the eikonal equation
in parallel to the Hamiltonian equation of motion.
Deviation from this geometrical optics is usually treated in terms 
of the diffraction theory. However, most of the analyses have been done in 
terms of the scalar diffraction theory, neglecting its vector 
nature.  However, the light has the degree of freedom of the polarization, 
which is represented by the spin $S=1$ parallel or anti-parallel to the 
wavevector $\bm{k}$. 
It is known that this spin produces the Berry phase when the light is 
guided by the optical fiber with torsion \cite{optfiber}. 
Also the adiabatic change 
of the polarization even without the change of $\bm{k}$ produces the
phase change called Pancharatnam phase \cite{Pancharatnam}. 
In this Letter, we will show that 
the trajectory or ray of light itself is affected by the Berry phase
\cite{Berry}, which leads to various nontrivial effects including the Hall 
effect.

For simplicity, we focus on an isotropic, nonmagnetic medium; 
the refractive index $n(\bm{r})$ is real and scalar, 
but is not spatially uniform.
Let us consider the wavefunction for the wave-packet with the wavevector 
centered at $\bm{k}_{c}$ and position centered at $\bm{r}_{c}$. 
Because of the conjugate relation of the position and wavevector, 
both of them inevitably have finite width of distribution. 
Therefore, the relative phase of the wavefunctions at 
$\bm{k}$ and $\bm{k} + d \bm{k}$ matters, 
which is the Berry connection or gauge field 
$ \bm{\Lambda}_{\bm{k}}$ defined by
$
[\bm{\Lambda}_{\bm{k}}]_{\lambda\lambda'}=
-i\bm{e}_{\lambda\bm{k}}^{\dagger}\bm{\nabla}_{\bm{k}}\bm{e}_{\lambda'\bm{k}}
$,
where $\bm{e}_{\lambda\bm{k}}$ is the polarization vector of 
$\lambda$-polarized photon, and $\lambda=\pm$ correspond
to the right- and left-circular polarization.
Here, it is noted that the SU(2) gauge field $\bm{\Lambda}_{\bm{k}}$ 
is $2 \times 2$ matrix.
The corresponding Berry curvature (BC) or field strength
is given by 
$
\bm{\Omega}_{\bm{k}} = 
\bm{\nabla}_{\bm{k}}\times\bm{\Lambda}_{\bm{k}}
+i\bm{\Lambda}_{\bm{k}}\times\bm{\Lambda}_{\bm{k}}
$.
Due to the masslessness of a photon,
it is diagonal in the basis of the right and left circular 
polarization as given by
$
\bm{\Omega}_{\bm{k}}=  \sigma_{3}\bm{k}/k^{3}
$
[a massive case will be discussed later].
It corresponds to the field radiated from the monopole with strength 
$\pm 1$ located at $\bm{k}=0$.
We pick up a correction to the geometric optics up to the linear order in 
$|\bm{\nabla}n(\bm{r}_{c})|/k_{c}$.
An equation of motion (EOM) is derived
including an effect of $\bm{\Lambda}_{\bm{k}}$
in the similar way as the Bloch waves of electrons \cite{Niu}.
%However, in the case of the electromagnetic field, 
%we must start from the second-quantized formalism
%in order to give the stochastic interpretation for quantum states.
%This is the most natural formalism, although the quantum mechanics 
%is not essential for derivation of the EOM; 
%it often happens that the classical Maxwell theory gives identical results 
%as those for the photon because the Maxwell equations are linear.
By considering the effective Lagrangian for the 
center coordinates $\bm{r}_{c}$ and $\bm{k}_{c}$, 
the variational principle leads to the following EOM \cite{Niu,Onoda}
\begin{eqnarray}
\dot{\bm{r}}_{c} & = & v(\bm{r}_{c}) \frac{\bm{k}_{c}}{k_{c}} 
+\dot{\bm{k}}_{c} \times(z_{c}|\bm{\Omega}_{\bm{k}_{c}}|z_{c}),
\label{eq:eom-r}\\
\dot{\bm{k}}_{c} & = & -[\bm{\nabla}v(\bm{r}_{c})]{k}_{c},
\label{eq:eom-k}\\
|\dot{z}_{c}) & = & -i\dot{\bm{k}}_{c} \cdot \bm{\Lambda}_{\bm{k}_{c}}|z_{c}),
\label{eq:eom-z}
\end{eqnarray}
where $v(\bm{r}) = 1/n(\bm{r})
$ is the velocity of light, 
and $|z) = \:^{t}[ z_{+},\ z_{-}]$ represents 
the polarization state. 
This is the central result of this Letter, from which the Hall effect 
of light is derived below in a straightforward way.

The limit of geometrical optics is obtained by neglecting
the terms containing $\bm{\Lambda}_{\bm{k}}$ or
$\bm{\Omega}_{\bm{k}}$, reproducing Fermat's principle. 
Let us study phenomena caused by 
$\bm{\Lambda}_{\bm{k}}$ and $\bm{\Omega}_{\bm{k}}$. 
First, the $\bm{\Lambda}_{\bm{k}}$ term
in Eq.~(\ref{eq:eom-z}) causes the phase shift 
by the directional change of 
the propagation discussed in ref.~\cite{optfiber}, as seen in the
following way.
The equation for $|\dot{z}_{c})$ gives the solution 
$|z_{c}^{\mathrm{out}})
=\:^{t}[e^{-i\Theta}z_{+}^{\mathrm{in}},e^{i\Theta}z_{-}^{\mathrm{in}}]$,
where $|z_{c}^{\mathrm{in}})
=\:^{t}[z_{+}^{\mathrm{in}}, z_{-}^{\mathrm{in}}]$
is the initial state of polarization.
$\Theta$ is the solid angle made by the trajectory of momentum: 
$\Theta=\oint d\bm{k}\cdot[\bm{\Lambda}_{\bm{k}}]_{++}
=\int_{S}d\bm{S}_{\bm{k}}\cdot[\bm{\Omega}_{\bm{k}}]_{++}$ 
where $d\bm{S}_{\bm{k}}$ is the surface 
element in $\bm{k}$-space and $S$ 
is a surface surrounded by the trajectory.
This is the phase shift by the directional change of
propagation \cite{optfiber}.

On the other hand, the second term in Eq.~(\ref{eq:eom-r}) 
induces a new effect found in this Letter; the trajectory of light is affected
by $\bm{\Omega}_{\bm{k}}$.  
We note that this term guarantees the conservation of 
total angular-momentum of photon
$
j_{z}  = [
\bm{r}_{c}\times\bm{k}_{c} + (z_{c}|\sigma_{3}|z_{c})
\frac{\bm{k}_{c}}{k_{c}}]_{z}
$,
assuming the rotational symmetry around the $z$-axis. From 
Eqs.~(\ref{eq:eom-r})-(\ref{eq:eom-z}), one can prove 
$d j_z/dt = 0$ when $\epsilon(\bm{r}) = \epsilon( \sqrt{x^2+y^2}, z)$.
Note that the orbital angular-momentum $\bm{r}_{c}\times \bm{k}_{c}$
is defined only when the position $\bm{r}_{c}$ is well-defined, 
which necessarily leads to the finite distribution of the wavevector 
and hence to the Berry phase. 
This argument applies to a light beam, whereas
it does not for the plane wave state. 

\begin{figure}[hbt]
\includegraphics[scale=0.25]{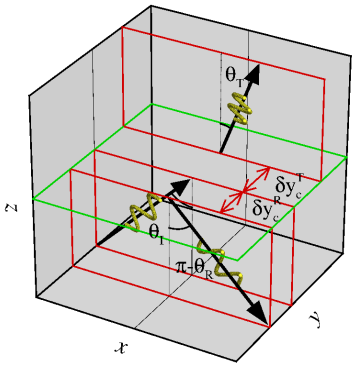}
\caption{
Transverse shift of light beams in the refraction 
and reflection at an interface.
}	
\label{fig:configuration}
\end{figure}

This Berry-phase term induces the Hall effect of light. 
Let us first consider reflection and refraction of light at a
flat interface between two media with different dielectric constants.
Let $z=0$ be the boundary between the two dielectric media with 
$n=v_{0}/v_{1}$ where $v_{0}$ and $v_{1}$ are the velocities of light 
in the media below and above the boundary,  
and $\bm{r}_{c}$ and $\bm{k}_{c}$ of the incident light 
are within the $y=0$ plane (Fig.~\ref{fig:configuration}).

The Snell's law tells us
$
\sin \theta_{T}/v_{1} = \sin \theta_{I}/v_{0}
$
for the transmitted light and 
$
\theta_{R} = \pi-\theta_{I}
$
for the reflected light,
where $\theta_{I,T,R}$ are the angle between the $z$-axis 
and $\bm{k}_{c}$ of the incident, transmitted and reflected lights.
This relation can be derived from Fermat's principle in geometrical optics. 
The question is ``what is the deviation from this Snell's law due to 
the Berry phase ?''
The answer is the transverse shifts of the centers of the reflected
and the transmitted wave-packets. This phenomenon could be regarded as
the ``Hall effect of light'', wherein the ``force'' to 
the light, i.e., the change of the dielectric constant, is 
perpendicular to the shift. 
The shift of the transmitted light can be easily calculated
from the EOM. 
Strictly speaking, the EOM can be applied only when the refractive 
index $n$ varies much slower than the wavelength;
thus the EOM is not reliable when the interface is sharp.
Nonetheless, it turns out to give the correct answer 
for the shift of the transmitted light even for 
the sharp interface, because the shift is 
governed  by the conservation law of 
the $z$-component $j_{z}$ of total angular-momentum.
We assume that $j_z$  is conserved in reflection and in refraction
separately: $j_{z}^{I}=j_{z}^{T}$, $j_{z}^{I}=j_{z}^{R}$, 
where the superscripts $I$, $T$ and $R$ stand for 
incident, transmitted and reflected lights, respectively.
This is required when the light is regarded as a quantized object, photon;
each photon is reflected or refracted with some probabilities. From 
this conservation of $j_{z}$,
the transverse shifts of transmitted and reflected wave-packets are
estimated as 
\begin{equation}
\delta y^{A}_{c}(\theta_{I}) = 
\frac{[(z_{c}^{A}|\sigma_{3}|z_{c}^{A})\cos\theta_{A}
-(z_{c}^{I}|\sigma_{3}|z_{c}^{I})\cos\theta_{I}]}{k_{c}^{I}\sin\theta_{I}},
\label{eq:delta_y}
\end{equation}
where $A=T$ or $R$, 
$k_{c}^{I}$ is the momentum of the incident light,
$|z_{c}^{I,T,R})$ stand for the polarizations of the incident, 
transmitted and reflected photons. 
We note  that $|z_{c}^{T(R)})$ is obtained from $|z_{c}^{I})$ and 
the transmittances (reflectances) of two eigenmodes with linear 
polarizations.

It has been already known that the totally reflected light beam 
undergoes shifts of position in the plane
of incidence (longitudinal) or normal to the plane of incidence (transverse).
The longitudinal shift was first studied by Goos and H{\"a}nchen 
\cite{Goos-Hanchen}.
The transverse shift was predicted by Fedorov \cite{Fedorov},
and have been observed experimentally by Imbert \cite{Imbert}, 
followed by a number of theoretical approaches 
\cite{totalreflection,energyflux}.
The transverse shift occurs not only in total reflection,
but also in partial reflection and refraction \cite{partialreflection}.
While there are several papers on theoretical predictions 
\cite{Fedoseev, partialreflection, angularmomentum}, 
they contain rather complicated
calculations and the issue remains still controversial.
The shift of the transmitted light has not been experimentally 
measured to the authors' knowledge. 
Our derivation based on the EOM is much clearer and simpler. 
Also the result is supported by the numerics described below,
showing a wide applicability of our EOM.

\begin{figure}[hbt]
\includegraphics[scale=0.25]{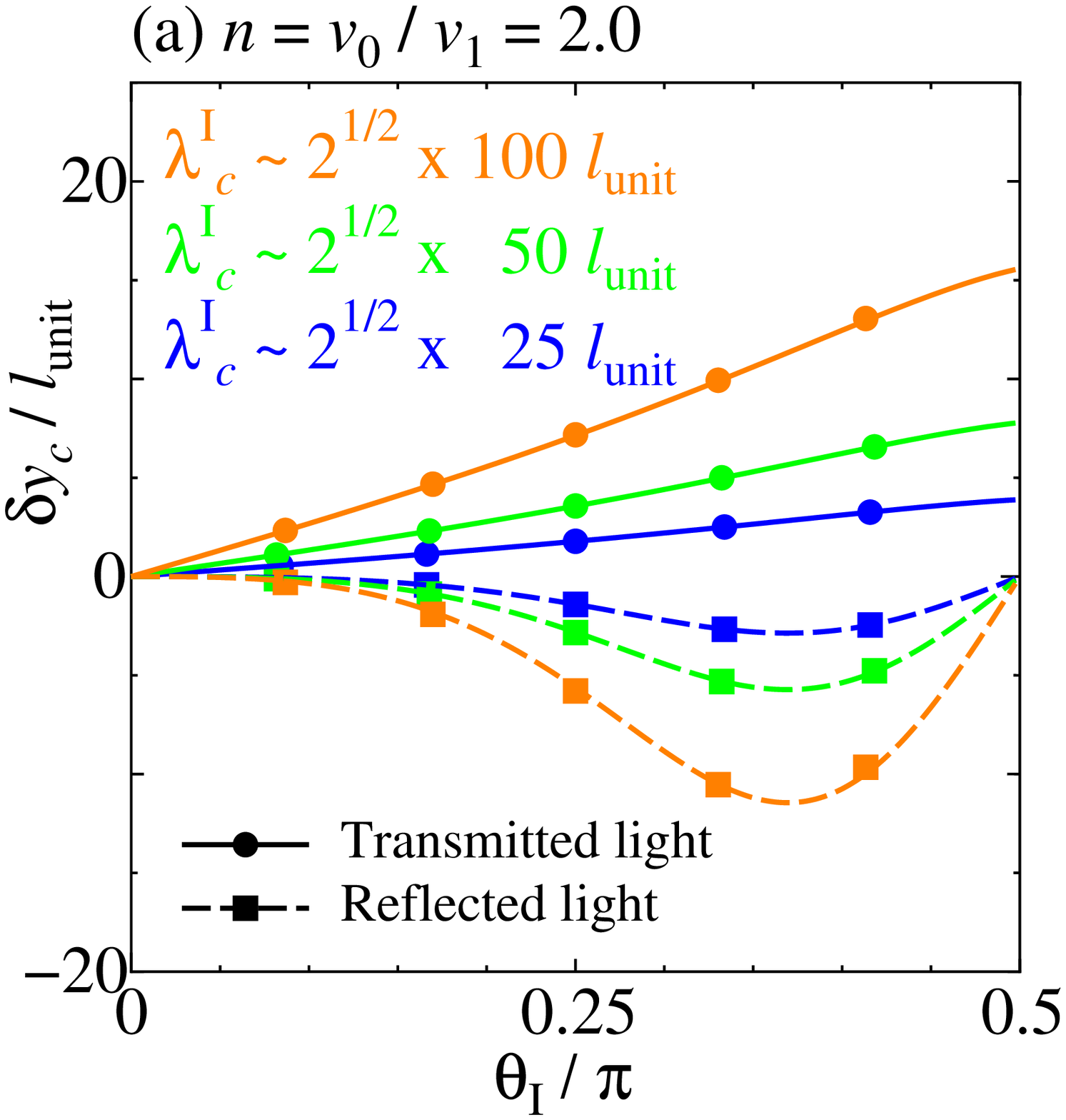}
\includegraphics[scale=0.25]{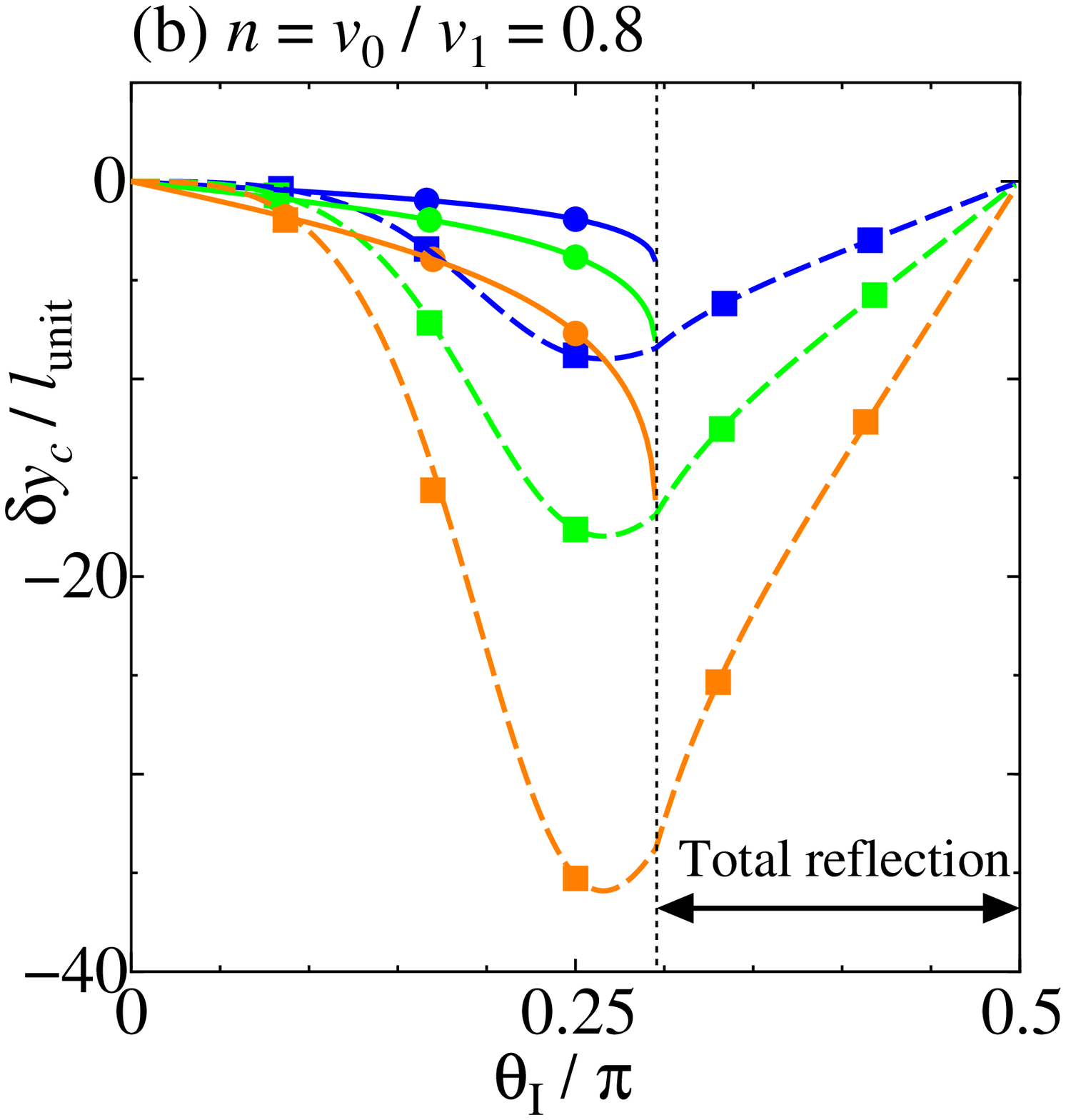}
\caption{
Transverse shift in the center of the transmitted and reflected 
wave-packets of light for (a) $n=2.0$ and (b) $n=0.8$
with the incident wave-packet being right-circularly polarized.
The solid and dashed lines for the analytic results 
Eq. (\ref{eq:delta_y}).
The filled circles and squares are the results of simulation.
$\lambda^{I}_{c} = 2\pi/k^{I}_{c}$ is the wavelength of incident wave-packet.
$l_{\mathrm{unit}}$ is an arbitrary length-scale.
The sign of the shifts is reversed for the left-circularly polarized light.
} 
\label{fig:shift}
\end{figure}

In order to test our prediction Eq.~(\ref{eq:delta_y}), 
we have numerically solved the Maxwell equation. 
It is found that both the reflected and transmitted wave-packets 
are shifted along the $y$ or $-y$ direction depending on the 
circular polarization, 
which perfectly  agrees with Eq.~(\ref{eq:delta_y})
as shown in Fig.~\ref{fig:shift}.
This phenomenon is analogous to the spin Hall effect \cite{Murakami,Sinova},
where the Berry phase from band structure 
gives rise to spin-dependent motion of electrons.

It is noted that the similar effect is expected also for massive
particle waves with intrinsic angular-momentum.
However, the effect is suppressed by the mass factor, $m$.
For example, for the spin-1/2 massive field,
the BC in helicity basis is given by
$
\bm{\Omega}_{\bm{k}}= 
\frac{1}{2E_{\bm{k}}^2}
[\frac{m}{E_{\bm{k}}}(\sigma_{1}\bm{e}_{\theta}
+\sigma_{2}\bm{e}_{\phi})+\sigma_{3}\bm{e}_{k}
],
$
where $E_{\bm{k}}=\sqrt{k^2+m^2}$ and 
$\bm{e}_{k,\theta,\phi}$ are the unit vectors
of the spherical coordinate system in $\bm{k}$-space.

The BC of light is proportional to $1/k^2$. 
Consequently, the transverse shift is a fraction of the wavelength, and 
is small for the visible light. 
In the following we propose a method
to enhance the Hall effect of light by use of photonic crystals
\cite{photonic-crystal}.
As is known, the periodic modulation of the dielectric constant leads to
formation of Bloch waves of light analogous to those of electron in solids. 
The degeneracy of right- and left-circularly polarized light 
is generally lifted.
Hence, the SU(2) gauge invariance breaks down to U(1) (Abelian).
This is similar to the anomalous Hall effect
in ferromagnetic materials \cite{Fang}. 
In photonic crystals, monopoles in $\bm{k}$-space can 
exist away from $\bm{k} = {\bf 0}$. 
The anomalous velocity is enhanced when $\bm{k}$ is near 
the monopole. 

\begin{figure}[hbt]
\includegraphics[scale=0.2]{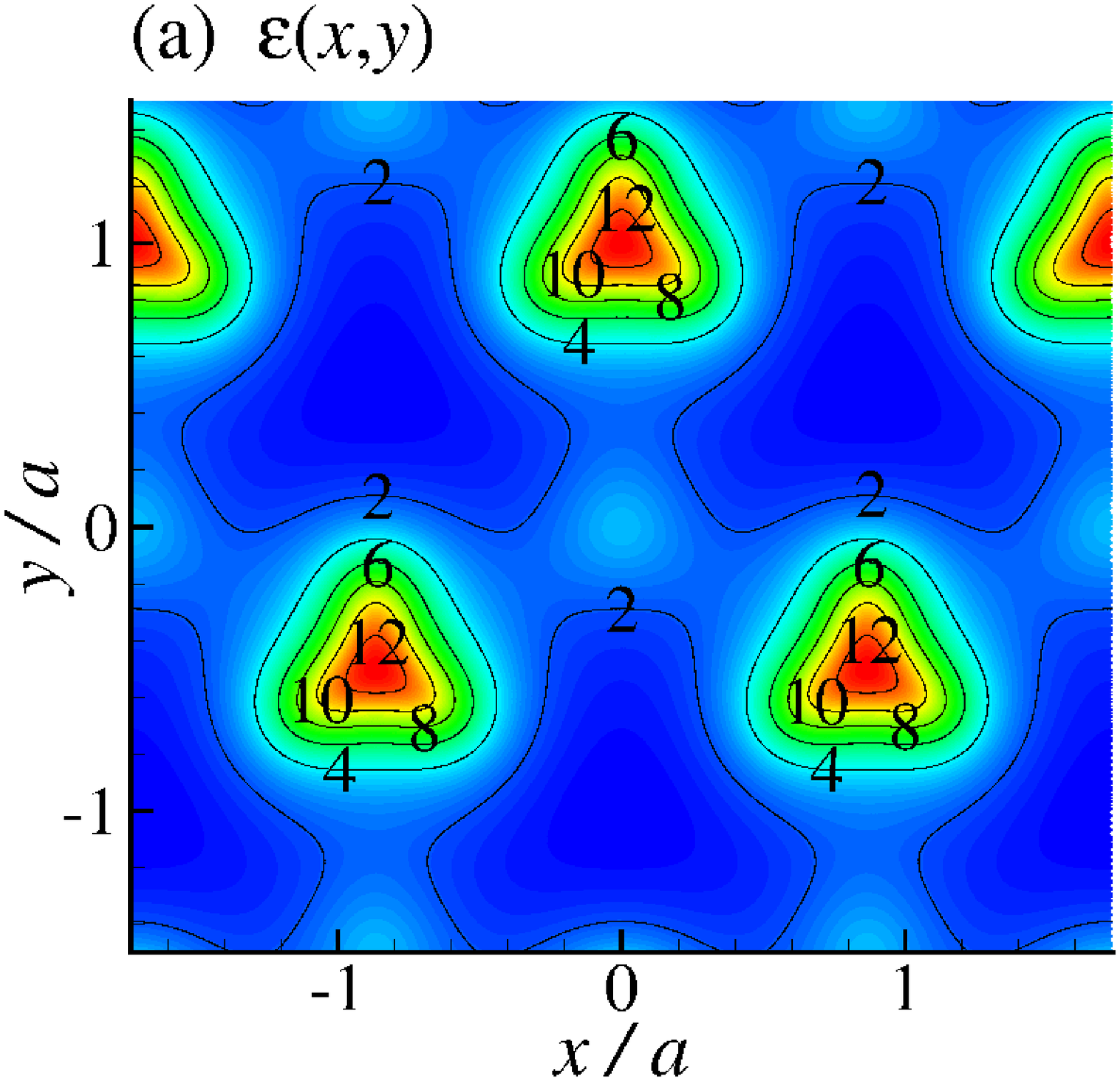}
\includegraphics[scale=0.25]{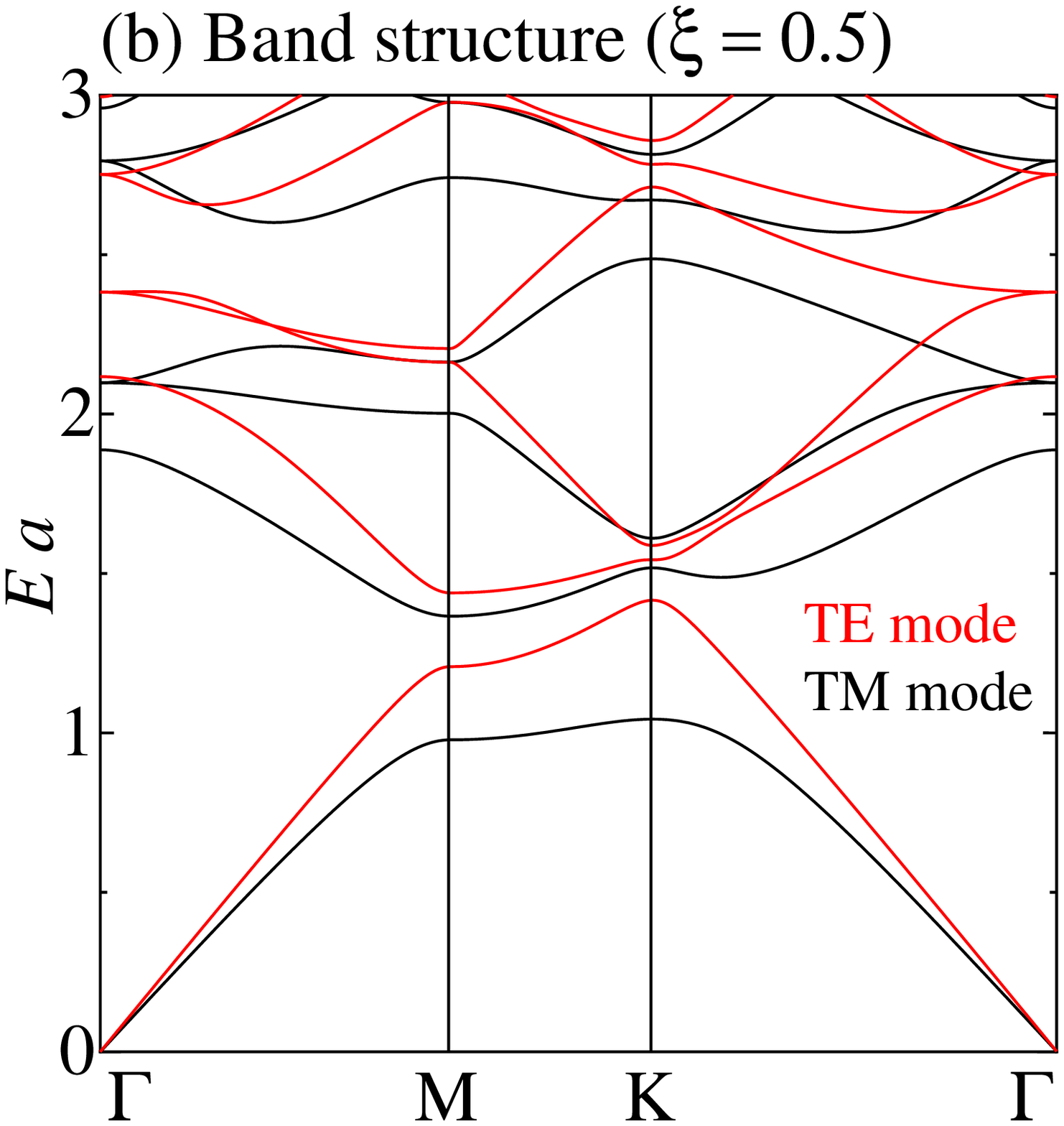}
\caption{
(a) Dielectric function 
and (b) band structure of the 2D photonic crystal.
The Brillouin zone is shown in Figs. \ref{fig:flux} and
\ref{fig:seom}(b)
}	
\label{fig:band}
\end{figure}
\begin{figure}[hbt]
\includegraphics[scale=0.25]{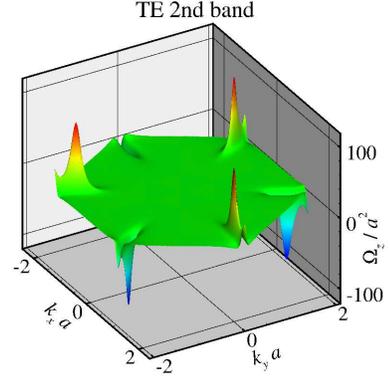}
\caption{Example of the
BC of the 2D photonic crystal.
It is prominent at the corners of the Brillouin zone.
}	
\label{fig:flux}
\end{figure}

For an explicit example, we consider a two-dimensional (2D) 
photonic crystal with the dielectric function $\epsilon(\bm{r})$,
$
\frac{1}{\epsilon(\bm{r})}
= \frac{4}{3(5+12|\xi|+8\xi^2)}
\sum_{i = 1}^{3}
[
(\xi-\cos(\bm{b}_{i}\cdot \bm{r}+\frac{2\pi}{3})
)^2
+(
\xi+\cos(\bm{b}_{i}\cdot\bm{r}-\frac{2\pi}{3})
)
],
$
where 
$\bm{b}_{1} = (\frac{2\pi\sqrt{3}}{3a}, -\frac{2\pi}{3a})$,
$\bm{b}_{2} = (0, \frac{4\pi}{3a})$,
and $\bm{b}_{3} = -\bm{b}_{1}-\bm{b}_{2}$.
The spatial distribution of $\epsilon(\bm{r})$ is 
shown in Fig.~\ref{fig:band}(a).
For simplicity, we shall focus on the case with $k_{z}=0$.
In other words, we shall consider wave-packets 
extended in $z$-direction, i.e. wave-ribbons.
The eigenmodes are classified into 
transverse magnetic (TM) modes and transverse electric (TE) modes. 
For the TM modes, the magnetic field is in the 
$xy$-plane, while for the TE modes it is in the $z$-direction.
The photonic bands are shown in Fig.~\ref{fig:band}(b).
The BC is along the 
$z$-direction in 2D, and its $z$-component is shown in Fig.~\ref{fig:flux}.
It is seen that $\bm{\Omega}_{\bm{k}}$ is strongly enhanced 
near the corners of the Brillouin zone, which can be interpreted as a 
2D cut of the monopole structure in the extended space including the 
parameters, e.g. $\xi$ in the present case.
[$\xi$ represents the degree of inversion-symmetry breaking.]

\begin{figure}[hbt]
\includegraphics[scale=0.25]{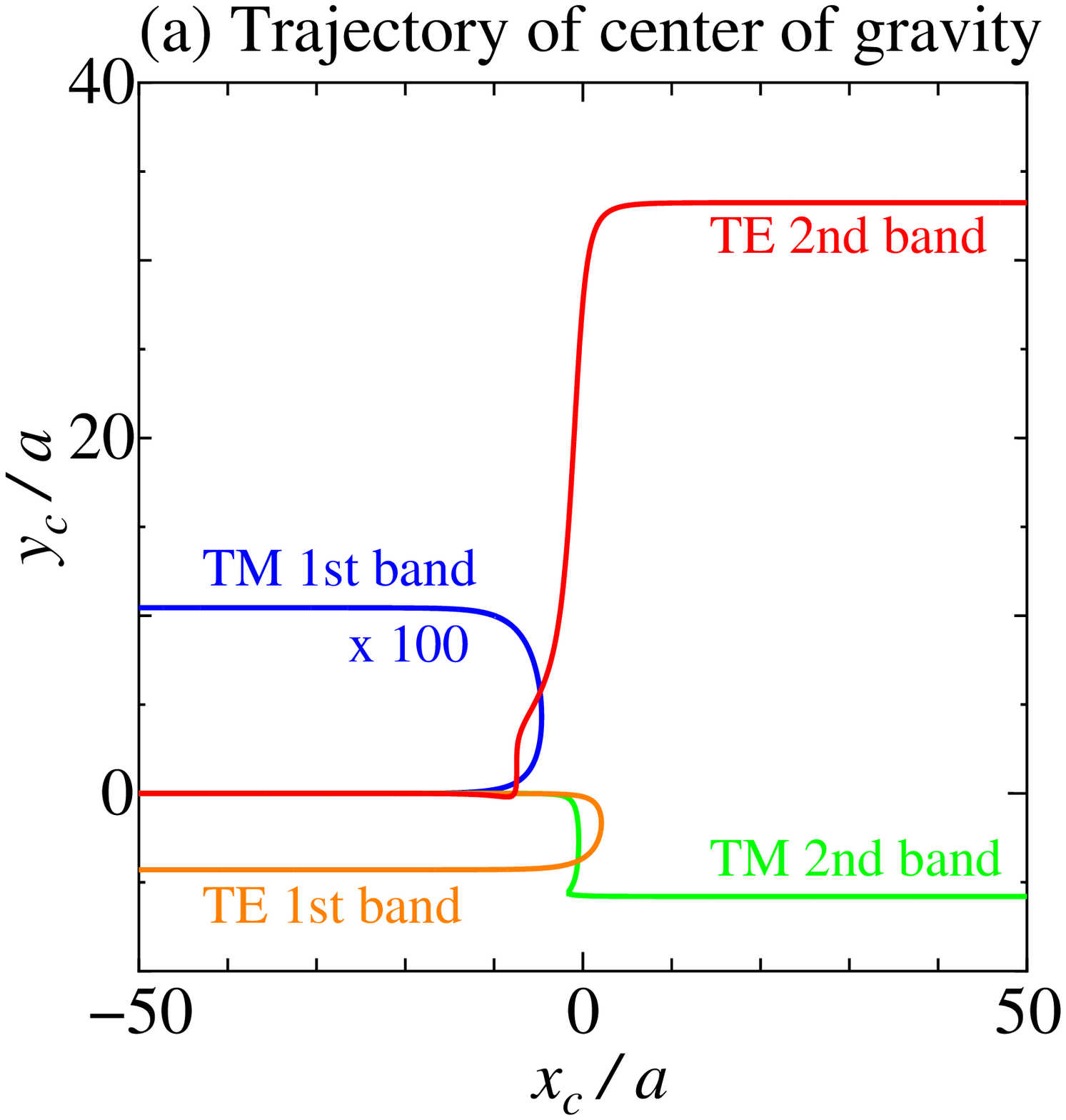}
\includegraphics[scale=0.3]{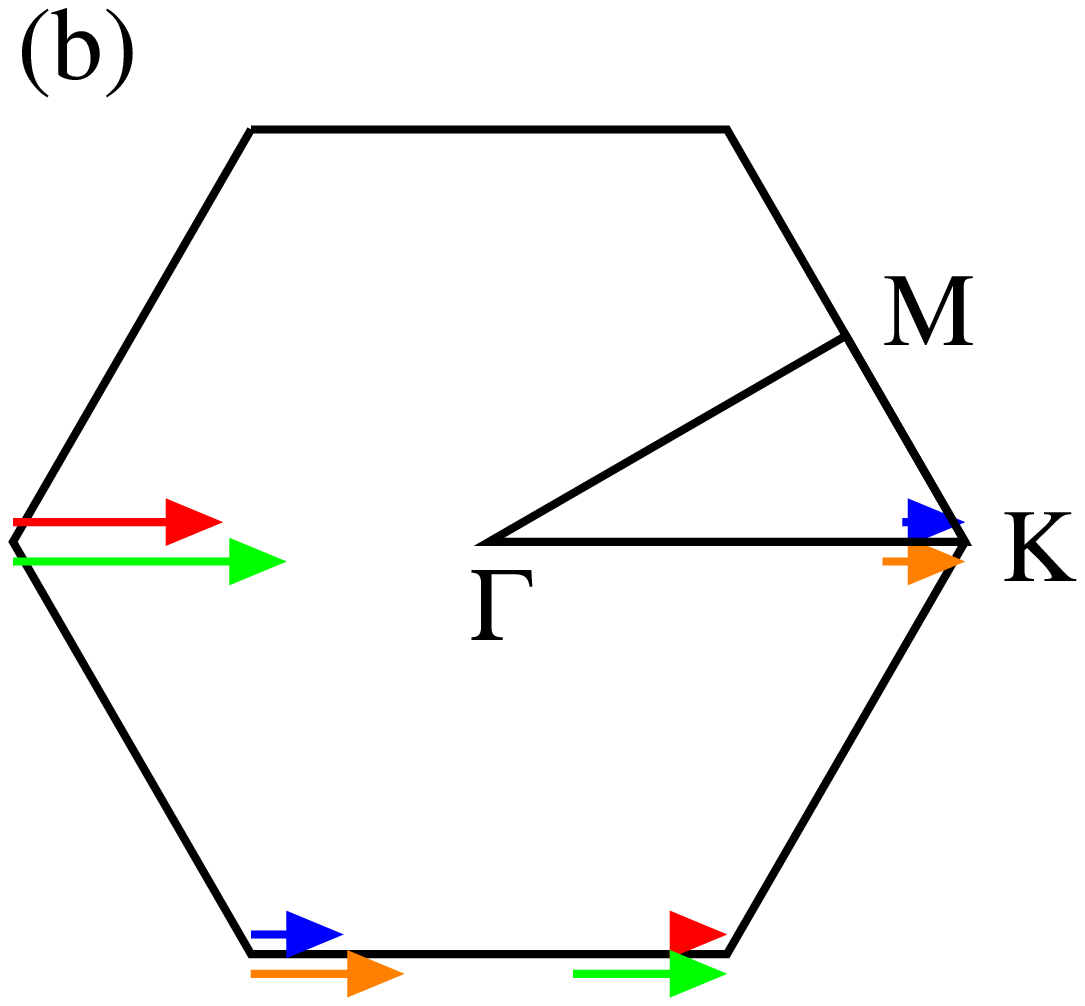}
\caption{
(a) Real-space trajectory of the center of the wave-packet.
(b) Momentum-space trajectory of the center of the wave-packet.
The arrows in the same colors in (a) and (b) refer to the same 
photonic modes.}	
\label{fig:seom}
\end{figure}

To induce the Hall effect of light in photonic crystals, 
we have to introduce 
a gradient of the refractive index, as seen in Eq.~(\ref{eq:eom-k}).
Therefore, we shall introduce a slow variation of the envelope of
the refractive index as
$n(\bm{r})\to n(\bm{r})/\gamma({x})$
where $1/\gamma(x)$ changes from 1 to 1.2 within the range of 
$w=10a$. We have solved the  EOM for the wave-packets in the photonic crystal,
$
\dot{\bm{r}}_{c} =
\gamma(x_{c})\bm{\nabla}_{\bm{k}_{c}}E_{n\bm{k}_{c}}
+\dot{\bm{k}}_{c}\times\bm{\Omega}_{n\bm{k}_{c}}
$,
$
\dot{\bm{k}}_{c} = -[\bm{\nabla}\gamma(x_{c})]
E_{n\bm{k}_{c}},
$
where $E_{n\bm{k}}$ and $\bm{\Omega}_{n\bm{k}}$ are the energy and 
the BC of the $n$-th band in the unperturbed crystal.
The EOM for $|z)$ brings about the phase shift for each mode.
We set the initial $\bm{k}_{c}$ near the corners of the Brillouin zone.
This is because the BC is large at the corners
and the effect by the anomalous velocity is expected to be prominent.
The obtained trajectories are shown in Fig.~\ref{fig:seom}.
It is found that the shift of $\bm{r}_{c}$
reaches to dozens of 
times the lattice constant especially for the TE second band.

We note that the BC is generally 
present in photonic crystals if the inversion-symmetry is broken.
Our EOM offers an easy method to 
calculate the trajectory of wave-packets for generic photonic crystals
\cite{note-slow}.
By contriving crystal structures, this effect can be enhanced 
considerably as follows.
The BC around the zone corners is determined 
mostly by the splitting, $2|\Delta|$, between the neighboring bands.
Suppose, for example, at $\bm{k}=\bm{k}_{0}$ another band 
comes very close in energy to the one considered.
The BC is approximately given by
$
\Omega_{z}\sim\frac{v^{2}\Delta}{
(\Delta^{2}+v^{2}|\bm{k}-\bm{k}_{0}|^{2})^{3/2}},
$
where $v$ is a nominal velocity of light.
Thus when the light traverses this $\bm{k}_{0}$ point, 
the shift is given by 
$
\delta y_{c}\sim 
\mathrm{sgn}[\nabla_{x_{c}}\gamma(x_c)]v/\Delta
$.
Therefore $|\delta y_{c}|$ is larger for smaller $|\Delta|$
as shown in Fig.~\ref{fig:seom}; the second gap is smaller and hence 
the wave-packet in the second band shows larger $|\delta y_{c}/a|$.

Let us discuss an effect of disorder in real photonic crystals.
It has been discussed that the photonic band with lower index 
is more robust against disorder
\cite{Asatryan}.
Hence, in order to see the Hall effect of light, it is better to use
the first band than the second band. 
While in Fig.~\ref{fig:seom} the shift of the 
first band is smaller than the second, it can be enhanced by 
reducing the gap between the two bands.
The first band is not affected 
appreciably when fluctuations of system parameters
are less than a few percent. 
For visible or near 
infrared light, this requires fabrication precision within
$\sim 50$nm, which is comparable to the current state of the art.
Experimentally, the band splitting $\Delta$ of the order of 
$10^{-5}$nm$^{-1}$ can be controlled
in the photonic crystal for the visible light \cite{Rosenberg}.
This corresponds to $\delta y_{c} \sim 1/\Delta \sim 100a$. Thus,
we can expect the shift as large as a hundred times the lattice constant $a$.

To conclude, we have derived the semiclassical 
equation of motion for the wave-packet
of light including the Berry curvature and resultant anomalous velocity.
This gives a natural generalization of geometrical optics including the 
wave nature of light, and leads to the Hall effect of light.
This also offers the natural framework to describe the ray of light 
in the photonic crystal without the inversion symmetry
where the Hall effect can be magnified tremendously.

The authors thank S.~C.~Zhang, A.~V.~Balatsky, K.~Miyano, and A.~Furusawa 
for fruitful discussions. M.~O. is supported by Domestic Research Fellowship
from Japan Society for the Promotion of Science. This work is financially 
supported by NAREGI Grant, Grant-in-Aids from the Ministry of Education,
Culture, Sports, Science and Technology of Japan, and by NEDO Grant.

{\it Note added} : 
A related argument on a spin transport of photons came to our attention 
after the submission of this work.
K. Yu. Bliokh and Yu. P. Bliokh, physics/0402110 (unpublished).


\begin{thebibliography}{99}
\bibitem{note-slow}
The EOM holds with good accuracy when the spatial variation of the 
envelope $\gamma(x)$ of the refractive index is slower than the 
wavelength. On the other hand,
the periodic variation of the refractive index in the 
photonic crystal need not to be slow.
\end{thebibliography}
\end{document}